\documentclass[seceq,supplement]{ptptex}
\usepackage{graphicx}

\title{%        
How many colors to color a random graph? \\
Cavity, Complexity, Stability and all that
}

\author{%       %Use \scshape  for the family name
Florent \textsc{Krz\c{a}ka{\l}a}%
}

\inst{%         %Affiliation, neglected when [addenda] or [errata]
Dipartimento di Fisica, INFM and SMC, Universit\`a di
Roma ``La Sapienza'', P.~A.~Moro 2, 00185 Roma, Italy}

\abst{We review recent progress on the statiscal physics study of the
problem of coloring random graphs with $q$ colors. We discuss the
existence of a threeshold at connectivity $c_q
=2q\log{q}-\log{q}-1+o(1)$ separting two phases which are respectivily
COL(orable) and UNCOL(orable) with $q$ colors; We also argue that the
so-called one-step replica symmetry breaking ansatz used to derive
these results give {\it exact} threshold values, and draw a general
phase diagram of the problem.}

\begin{document}

\maketitle

Since the first observation in 1852 by Francis Guthrie that any planar
map could be colored with only $4$ colors, graph coloring has grown to
become an important problem both in combinatorial
mathematics\cite{GaJo} and in statistical physics\cite{WU}. Given a
graph, or a lattice, and given a number $q$ of colors, it consists in
assigning a color to each vertex such that {\it no} edge has two
equally colored end vertices. When defined on random
graphs\cite{Erdos_Renyi}, the problem turns out to be NP-Complete and
to display an interesting phase transitions at the so-called
$q$-COL/UNCOL connectivity $c_q$: graphs of average connectivity
$c<c_q$ do have proper $q$-colorings with high probability
(approaching one for graph size $N\to\infty$), whereas graphs of
higher connectivity require more than $q$ colors. Here we review the
recent progress on the statistical physics approach to the
characterization of the phase diagram of this problem.

A very first approximation for physicists working in disordered
systems is the so-called annealed computation (the first moment method
in computer science). Take two connected vertices: the probability
that they share the same color for a random assignment is $1/q$, hence
they have different colors with probability $1-1/q$. A crude estimate
of the probability that a random configuration colors a graph of
average connectivity $c$ is easily obtained: there are $cN/2$ links
and each of them has a probability $1-1/q$ to be satisfied, therefore
the number of COL configurations is
\begin{equation}
{\cal{N}}(c) \propto q^N \left( 1-\frac{1}{q} \right)^{cN/2} \propto
e^{N\Sigma(c)}, ~~~~ {\text where } ~~ \Sigma(c) = \log{q} +
\frac{c}{2} \log{(1-1/q)}.
\label{eq}
\end{equation}
It is straightforward to deduce from the preceding formula the
existence of a critical connectivity $c_q \approx 2q\ln q -\ln q$. For
$c>c_q$, $\Sigma(c)<0$ and the number of COL assignments is vanishing
exponentially with the size of the graph while for $c<c_q$,
$\Sigma(c)>0$ and therefore there is an exponentially huge number of
COL assignments: the COL/UNCOL transition is easily seen already at
the annealed level.  Such considerations are far from being only
hand-waving; in fact, similar computations, using the first and second
moment methods, allow to {\it rigorously} show that\cite{Lu} $2q\ln
q - \ln q -1 + o(1) \geq c_q \geq 2q \ln q -2 \ln q + o(1)$. To go
beyond these inequalities, we turn toward the use of more complex
statistical physics tools.

\begin{figure}
       \centerline{\includegraphics[width=13cm]{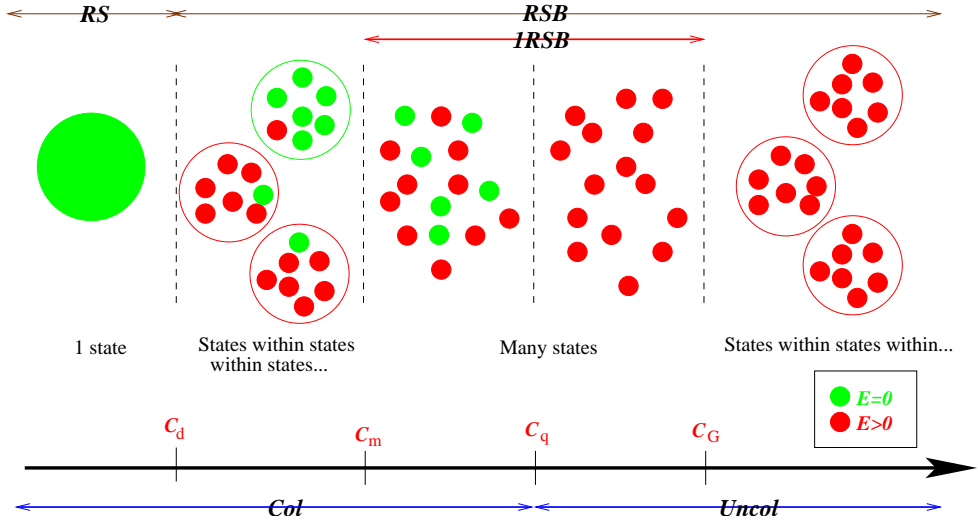}}
   \caption{Generic phase diagram of the coloring problem on random
   graphs. The COL/UNCOL transition happens at $c_q$, but the phase
   diagram is more complex and the system also undergoes different
   replica symmetry breaking (see text).}
   \label{fig:1}
\end{figure}
It is quite immediate, for a statistical physicist, to realize that
the coloring problem is equivalent to knowing if the energy of the
ground-state of an anti-ferromagnetic Potts model\cite{PottsGlass} on
a random graph is zero (COL) or not (UNCOL). Inspired by spin glass
theory\cite{BOOK}, one can use the replica/cavity approach to
analytically study the problem, following the seminal work
of \citen{Monasson}. In this particular context of constraint
satisfaction problems, the zero temperature cavity method is
particularly well suited\cite{BOOK} and thanks to many recent
developments\cite{Cavity,MeZe}, precise and detailed studies have
been made possible\cite{Saad,MeZe,Coloring1,Coloring2}. Although the
method is not fully rigorous, the self-consistency of the main
underlying hypothesis has been checked in different optimization
problems\cite{MZ2,MPR,me_col} and therefore all the features of the
solution derived by the statistical physics approach we will review
here are conjectured to be exact results, not approximations. Of
course, it is of first importance to develop rigorous mathematical
approaches in order to confirm them\cite{Silvio}.

The analytical results on the phase diagram are summarized on
Fig.\ref{fig:1}, it is very similar to the one first observed for
other optimization problems such as the
K-Satisfiability\cite{MZ2}. Let us discuss it for the particular case
of the $3-$coloring (we refer the readers to \citen{Coloring1,Coloring2,me_col} for more details).  When one
varies the connectivity $c$, there actually exist many distinct
phases, separated by thresholds connectivity $c_d$, $c_m$, $c_q$ and
$c_{SP}$. The most important point is of course the critical COL/UNCOL
transition that happens at $c_q =4.69$. It separates the COL phases at
$c<c_q$ from the UNCOL phase at $c>c_q$.  But in the COL region, there
actually exist distinct phases that differ by the structure of their
phase space. First, For $c<c_d\simeq 4.42$, the set of COL assignments
builds one cluster which is basically connected and from one single
valley in the phase space. This phase is called the EASY-COL phase,
for it is generally quite easy for any algorithm to find a COL
assignment in this phase.

On the other hand, for $c>c_d$, this phase space becomes disconnected
and the COL assignments are grouped into many clusters: this is the
phenomenon of Replica Symmetry Breaking (RSB), familiar to anyone who
had practiced mean field spin glass theory. This phase space is
characterized by the presence of many non-ergodic valleys and it is
now impossible for any physical dynamics to get from one cluster to
another. Moreover, the phase space develops many metastable states at
higher energies (corresponding to UNCOL configurations) and any too
simple algorithm, such as steepest descent or simulated annealing,
would get trapped into some these metastable states in a same way
aging systems undergo a glass transition; the COL region in this phase
is thus said to be HARD-COL. For $c>c_q$, there are no more COL
solutions and all valleys are UNCOL. In the whole RSB phase, the
number of such valleys is growing exponentially with the size of the
problem and the cavity method allows the computation of the logarithm
of this number in a very similar way one computed the annealed
logarithm number of COL solutions $\Sigma(c)$ in eq.(\ref{eq}). This
quantity, which is called the {\it complexity} is a central element
within the cavity method. All these results where first obtained by
Zecchina and collaborators\cite{Coloring1,Coloring2}.

This complex structure of the phase space is however very hard to
study. In fact it is only when it is not too complex, i.e. when
configurations simply group into valleys ---the so-called one-step
Replica Symmetry Breaking (1RSB) ansatz--- that one is able to solve
the equations. However the situation is generically much more complex:
we {\it do} have valleys within valleys within valleys etc\ldots which
obviously make the whole approach very difficult! In fact, most
computations are treating the phase space {\it as if} it was only
1RSB, neglecting the effect of valleys within valleys. Recently, by
studying more precisely the structure of this phase space, we
shown\cite{me_col} that, luckily enough, there is a zone in this
complex RSB phase which is indeed 1RSB; for the 3-coloring, it happens
for $c_{m} \approx 4.51 <c<c_{G} \approx 5.08 $. Even more luckily, it
turns out that $c_q$, the COL/UNCOL threshold connectivity, is {\it
precisely} inside this zone\cite{me_col}, and that we are thus able to
compute it without doing {\it any} approximation. In other word, the
original computation of\cite{Coloring1} was made in a {\it stable
1RSB} zone, and is therefore valid. This knowledge of the phase space
allows us to draw a more complete and quite generic phase diagram
(fig.\ref{fig:1}) where we show all the different phases and
transitions (COL/UNCOL, RSB/1RSB\ldots) that the system undergoes
while varying connectivity.

\begin{figure}
\centerline{\includegraphics[]{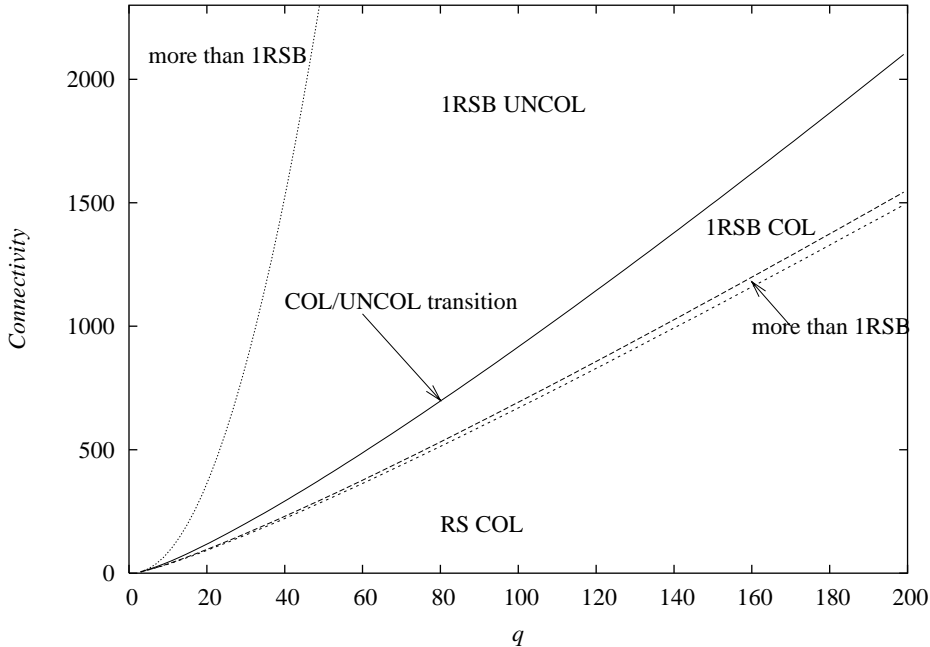}}
   \caption{Phase diagram of the $q-$coloring problem on regular fixed
connectivity random graphs.}
   \label{fig:2}
   \end{figure}
It turns out that the feature of this phase diagram are the same for
every number of colors $q \geq 3$\cite{me_col} (as well as for other
satisfaction problem\cite{MPR,MZ2}). For instance, in the specific
case of random graph with {\it fixed} connectivity (where graphs are
still random but constructed in such a way that each vertexes have the
same connectivity) one can derived analytically all critical
connectivities for any number $q$ of colors. We illustrate these
computations in Fig.\ref{fig:2}, where the different phases and
transition discussed here are clearly seen. It can be checked directly
that the critical COL/UNCOL transition always happens {\it inside} the
1RSB stable zone. Within this approach, we also shown\cite{me_col}
that the value of the COL/UNCOL threshold is asymptotically $c_q = 2q
\log{q} - \log{q} -1 + o(1)$\cite{me_col}, which agrees {\it
perfectly} with mathematical bounds mentioned before.

To conclude, the statistical physics approach, via the cavity method,
of the coloring problem turns out to be rather fruitful. Not only it
is consistent with independently established rigorous mathematical
results, but it also allows for calculation and determination of the
phase diagram and for a sharper, though not rigorous, determination of
threshold values.

I would like to warmly thank A. Pagnani and M. Weigt for the collaboration
that led to the a large part of the results presented here. I benefit from
many discussions with M.~M\'ezard, F.~Ricci-tersinghi and R.~Zecchina.  I
acknowledge support from European Community's Human Potential program under
contract HPRN-CT-2002-00319 (STIPCO), the ISI Foundation and the EXYSTENCE
Network.


\begin{thebibliography}{99}

\bibitem{GaJo} M. R. Garey and D. S. Johnson, \textit{Computers and
    intractability} (Freeman, New York, 1979).

\bibitem{WU} F.Y. Wu, Rev. Mod. Phys. {\bf 54}, 235 (1982).

\bibitem{Erdos_Renyi} P. Erd\"os and A. R\'enyi, Publ. Math.
  (Debrecen) {\bf 6}, 290 (1959).
  
\bibitem{Lu} T. Luczak, Combinatorica {\bf 11}, 45 (1991). D. Achlioptas and A. Naor, preprint (2004).

\bibitem{PottsGlass} D. J. Gross, I. Kanter and H. Sompolinsky,
Phys. Rev. Lett. {\bf 55}, 304-307 (1985).

\bibitem{BOOK}
M. M{\'e}zard, G. Parisi, and M.~A. Virasoro, {\em Spin-Glass Theory and
  Beyond}, Vol.~9 of {\em Lecture Notes in Physics} (World Scientific,
  Singapore, 1987).

\bibitem{Monasson} R. Monasson, R. Zecchina, S. Kirkpatrick {\it et
al.} Nature (London), {\bf 400}, 133 (1999).

\bibitem{Cavity} M. M\'ezard, G. Parisi, and R. Zecchina, Science {\bf
297}, 812 (2002).

\bibitem{MeZe} M. M{\'e}zard and R. Zecchina, Phys. Rev. E, {\bf 66},
  056126, (2002).

\bibitem{Saad} J. van Mourik and D. Saad, Phys. Rev. E {\bf 66},
056120 (2002).

\bibitem{Coloring1} R. Mulet, A. Pagnani, M. Weigt and R. Zecchina,
Phys. Rev. Lett. {\bf 89}, 268701 (2002)

\bibitem{Coloring2} A. Braunstein, R. Mulet, A. Pagnani {\it el al.},
Phys. Rev. E {\bf 68}, 036702 (2003).

\bibitem{MZ2} S. Mertens M. M{\'e}zard and R. Zecchina, CC/0309020

\bibitem {MPR} A. Montanari, G. Parisi and F. Ricci-Tersenghi,
J. Phys. A {\bf 37}, 2073 (2004)
 
\bibitem{me_col} F.~Krzakala, A.~Pagnani and M.~Weigt, \PRE{70,2004,046705}.

\bibitem{Silvio} S.~Franz and M.~Leone, J. Stat. Phys. {\bf 111} (2003) 535

\end{thebibliography}
\end{document}